\documentclass[%
reprint,
superscriptaddress,
amsmath,amssymb,
aps,
prb,
]{revtex4-2}

\usepackage{graphicx}
\usepackage{dcolumn}
\usepackage{bm}

\usepackage[utf8]{inputenc}
\usepackage[T1]{fontenc}
\usepackage{booktabs, array, mathptmx, float, tabularx, booktabs, lipsum, amsmath,multirow}
\usepackage{siunitx, xcolor}
\usepackage[version=4]{mhchem}

\usepackage{times}
\usepackage{amsmath}
\usepackage{mathptmx}
\DeclareMathAlphabet{\mathcal}{OMS}{cmsy}{m}{n}
\DeclareSymbolFont{largesymbols}{OMX}{cmex}{m}{n}
\usepackage{mathrsfs}
\usepackage{epstopdf}
\usepackage[colorlinks=true, letterpaper=ture, pdfstartview=FitV, linkcolor=blue, citecolor=blue, urlcolor=blue]{hyperref}
\usepackage{dcolumn}

\begin{document}

	\title{Suppression of the Skyrmion Hall Effect in Synthetic Ferrimagnets with Gradient Magnetization}%
	
	\author{Lan Bo}%
	\affiliation{Key Laboratory for Anisotropy and Texture of Materials (MOE), School of Materials Science and Engineering, Northeastern University, Shenyang 110819, China}
	\affiliation{Department of Applied Physics, Waseda University, Okubo, Shinjuku-ku, Tokyo 169-8555, Japan}
	\author{Xichao Zhang}%
	\affiliation{Department of Applied Physics, Waseda University, Okubo, Shinjuku-ku, Tokyo 169-8555, Japan}
	\author{Masahito Mochizuki}%
    \email[Corresponding E-mail: ]{masa\_mochizuki@waseda.jp}
	\affiliation{Department of Applied Physics, Waseda University, Okubo, Shinjuku-ku, Tokyo 169-8555, Japan}
	\author{Xuefeng Zhang}%
	\email[Corresponding E-mail: ]{zhang@hdu.edu.cn}
	\affiliation{Key Laboratory for Anisotropy and Texture of Materials (MOE), School of Materials Science and Engineering, Northeastern University, Shenyang 110819, China}
	\affiliation{Institute of Advanced Magnetic Materials, College of Materials and Environmental Engineering, Hangzhou Dianzi University, Hangzhou 310012, China}
	
	\date{\today}%

	\begin{abstract}
		Magnetic skyrmions are promising building blocks for future spintronic devices. However, the skyrmion Hall effect (SkHE) remains an obstacle for practical applications based on the in-line transport of skyrmions. Here, we numerically study the static properties and current-driven dynamics of synthetic ferrimagnetic skyrmions. Inspired by graded-index magnonics, we introduce a linear gradient of saturation magnetization ($M_{\rm s}$) in the skyrmion-hosting sample, which effectively modulates the skyrmion Hall angle and suppresses the SkHE. Micromagnetic simulations reveal that ferrimagnetic skyrmions could exhibit greater susceptibility to the variation of $M_{\rm s}$ as compared to their ferromagnetic counterparts. The Thiele analysis is also applied to support the simulation results, which elucidates that the $M_{\rm s}$ gradient dynamically modifies the intrinsic normalized size of skyrmions, consequently impacting the SkHE. Our results pave the way to the graded-index skyrmionics, which offers novel insights for designing ferrimagnet-based skyrmionic devices.
		
	\end{abstract}
	
	\maketitle
	
	\section{\label{sec:1}Introduction}
	
	Magnetic skyrmions are topologically nontrivial spin textures \cite{roessler2006spontaneous,nagaosa2013topological,mochizuki2015dynamical,jiang2017skyrmions,kanazawa2017noncentrosymmetric} characterized by an integer topological charge number $Q=(4\pi)^{-1}\iint{\bf m}\cdot \left({\partial_x {\bf m}} \times {\partial_y {\bf m}}\right) \, d^2r$ \cite{nagaosa2013topological}, and hold great potential as nonvolatile information carriers in next-generation classical and quantum spintronic devices \cite{finocchio2016magnetic,kang2016skyrmion,fert2017magnetic,everschor2018perspective,tokura2020magnetic,back20202020,zhang2020skyrmion,bogdanov2020physical,dieny2020opportunities,bo2022micromagnetic,psaroudaki2021skyrmion,psaroudaki2023skyrmion,xia2023universal}. Although the skyrmionics field has predominantly focused on skyrmions in ferromagnetic (FM) materials since their initial experimental observation \cite{yu2010real}, challenges may arise from the skyrmion Hall effect (SkHE) when driving FM skyrmions by spin currents \cite{zang2011dynamics,iwasaki2013current,iwasaki2013universal,jiang2017direct,litzius2017skyrmion,chen2017skyrmion,tomasello2018micromagnetic}. This effect leads to a transverse motion due to the $Q$-dependent Magnus force \cite{nagaosa2013topological}, often resulting in undesired accumulation and annihilation of skyrmions at device edges \cite{purnama2015guided,zhang2015skyrmion}. A promising avenue to avoid the SkHE involves the use of antiferromagnetic (AFM) skyrmions \cite{rosales2015three,barker2016static,zhang2016thermally,zhang2016magnetic,tomasello2017performance,duine2018synthetic,moriyama2018spin,dohi2019formation,xia2019current,legrand2020room, salimath2020controlling,gao2020fractional,xia2021current,mohylna2022spontaneous,aldarawsheh2022emergence,aldarawsheh2023spin,barker2023breathing,barker2024phase,aldarawsheh2024current}. While AFM skyrmions exhibit diminished SkHE \cite{zhang2016magnetic} and potential for fast spin dynamics \cite{jungwirth2016antiferromagnetic}, their insensitivity to external stimuli, such as the magnetic field, limits the manipulation and detection \cite{weissenhofer2023temperature}.

	A balanced solution is to construct the ferrimagnetic (FiM) skyrmions \cite{weissenhofer2023temperature,kim2017self,kim2019tunable,woo2018current,hirata2019vanishing,brandao2019evolution,mandru2020coexistence,yildirim2022tuning,berges2022size,wang2022topological}, which could combine advantages from both FM and AFM materials \cite{weissenhofer2023temperature}. It has been demonstrated that FiM skyrmions exhibit a reduced small but non-zero SkHE \cite{woo2018current,wang2022topological}. An example of FiM skyrmions can be found in the synthetic FiM system based on rare-earth-transition-metal (such as Gd-Co) alloys \cite{woo2018current,hirata2019vanishing,brandao2019evolution,berges2022size} or multilayers \cite{wang2022topological,blasing2018exchange,mullermodelling,li2023ultrafast}, which has garnered significant interest in recent years. In such a system, the two FM components are antiferromagnetically coupled with a bilinear surface exchange interaction \cite{das2023bilayer}.
	
	Following the principles inspired by graded-index optics \cite{shen2024topologically}, researchers have proposed a continuous modulation of magnetic parameters in spintronics \cite{davies2015graded,davies2015towards,davies2017mapping}. Although previous studies have delved into scenarios involving magnetic anisotropy gradients \cite{tomasello2018chiral,shen2018dynamics,de2023skyrmion} and  Dzyaloshinskii–Moriya interaction (DMI) gradients \cite{udalov2021magnetic,sapozhnikov2022zigzag,gorshkov2022dmi}to control skyrmion dynamics, models with saturation magnetization ($M_{\rm s}$) gradients have only been extensively investigated in the context of magnonics \cite{vogel2018control,borys2019scattering,gallardo2019spin,mieszczak2020anomalous,borys2021unidirectional}, but have not yet been explored in the study of skyrmion dynamics. The $M_{\rm s}$ of FiM materials exhibit higher sensitivity to temperature changes compared to FM materials, which provides a natural advantage in constructing $M_{\rm s}$ gradients through thermal landscapes \cite{kolokoltsev2012hot,vogel2015optically}.
	
	In light of the current fervor in both FiM skyrmions and graded-index spintronics, in this work, we numerically investigate the skyrmion dynamics in synthetic FiM with a $M_{\rm s}$ gradient. It is demonstrated that introduction of the $M_{\rm s}$ gradient can effectively regulate the skyrmion Hall angle. A theoretical analysis further reveals that the regulation is attributed to the variation of normalized skyrmion radius. The paper is organized as follows. In Sec. \ref{sec:2}, we describe the micromagnetic and analytical model. In Sec. \ref{sec:3}, we show and discuss the results of the numerical studies, which are summarized in Sec. \ref{sec:4}. Specifically, in Sec. \ref{sec:31}, we verify how static skyrmion configuration varies with changes in material parameters, preparing for subsequent dynamic studies. In Sec. \ref{sec:32}, two types of current injection geometries are compared, and an overall trajectory of skyrmion motion induced by spin-orbit torque (SOT) is provided. The details of skyrmion dynamics are presented, demonstrating an effective suppression of SkHE in the presence of the $M_{\rm s}$ gradient. In Sec. \ref{sec:33}, we analyze the physical mechanisms behind the simulation results based on the Thiele approach. 
	
	\section{\label{sec:2}Model and methodology}
	The system considered in our simulations is a synthetic FiM Gd/Co/Pt multilayer \cite{wang2022topological,blasing2018exchange,mullermodelling,li2023ultrafast}. As depicted in Figs.~\ref{1} (a) and (b), the magnetization in Gd and Co layers are oppositely aligned due to the AFM coupling, leading to the opposite polarities of skyrmions in the two layers. This kind of magnetization configuration is known as the exchange-coupled bilayer-skyrmion \cite{xia2019current,xia2021current,wang2022topological}. As shown in Fig.~\ref{1} (c), a nanoplate geometry with dimensions of $300\times200$ nm is considered to investigate skyrmion dynamics. This nanoplate includes regions at left and right sides with constant $M_{\rm s}$ values (i.e. $M_{\rm s}$ and $M_{\rm s}^*$), each spanning a width of 50 nm. In the central square region of $200\times200$ nm, spatially varying $M_{\rm s}$ values with a linear gradient $\varDelta M_{\rm s}$ are introduced along the $x$-axis. Periodic boundary conditions are imposed in the $y$-direction to remove the potential boundary effect. With a single isolated skyrmion initially placed at the left side, a spin-polarized current is applied to drive this skyrmion into motion towards the $+x$ direction.
	
	\begin{figure}[t]
		\includegraphics[width=1\linewidth]{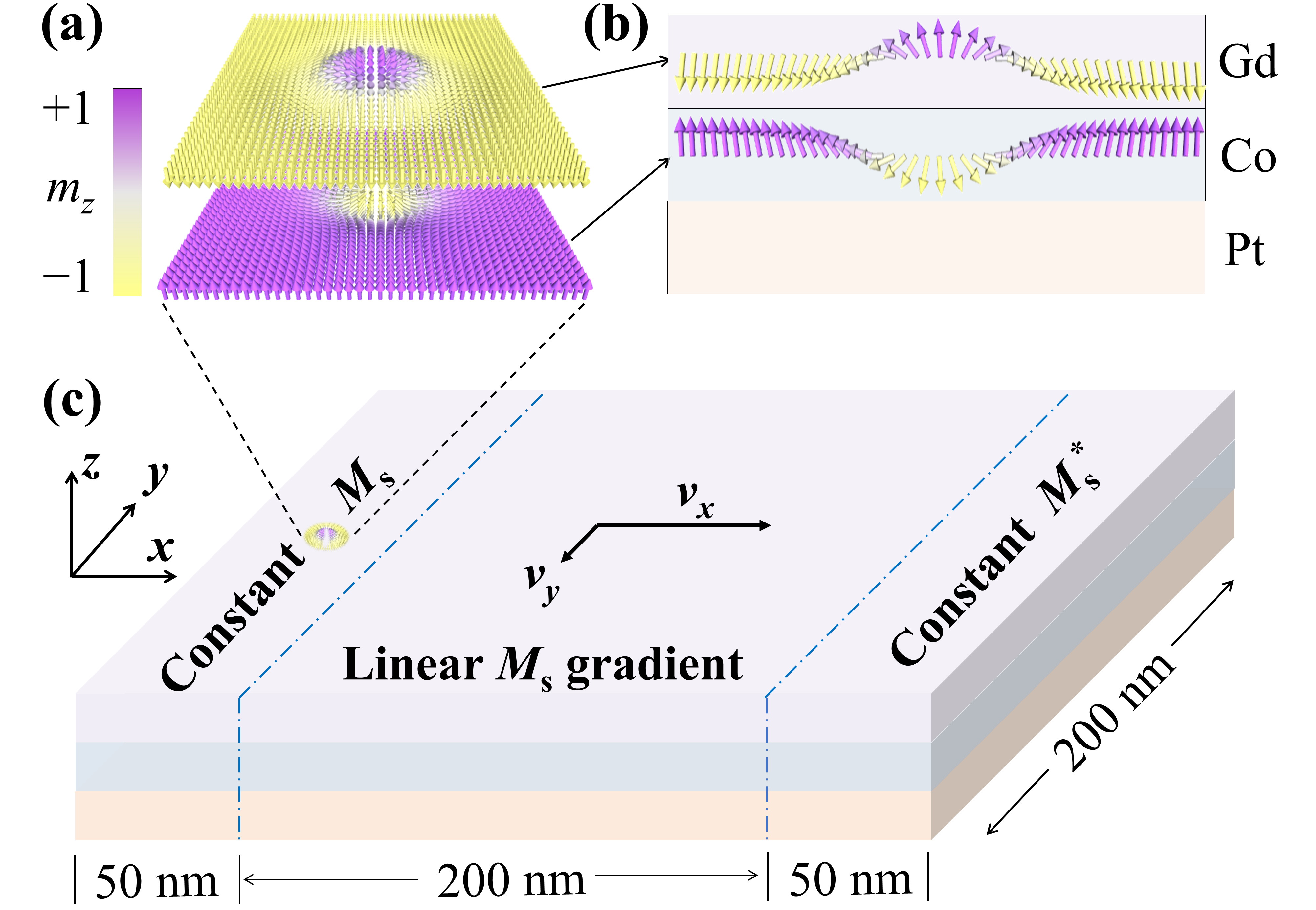}
		\caption{(a) Schematic configuration of a ferrimagnetically exchange-coupled bilayer skyrmion. Arrows represent the magnetization, and the color represents their $z$-component $m_z$. (b) Cross-sectional view of the Gd/Co/Pt heterojunction. Only the magnetization in free layers are explicitly modeled. (c) Schematic diagram of the geometry for skyrmion dynamics. The central region has a linear magnetization gradient $\varDelta M_{\rm s}$, and the left and right regions have constant $M_{\rm s}$.}\label{1}  
	\end{figure}
	
	The total Hamiltonian $H$ of the system includes intralayer terms $H_{\rm intra}^{L}$ for respective layers ($L=$\ Co, Gd) and an interlayer term $H_{\rm inter}$, which is given by
	\begin{align}
		&H=H_{\rm intra}^{\rm Co}+H_{\rm intra}^{\rm Gd}+H_{\rm inter}.\label{1}				
	\end{align}
	The intralayer Hamiltonian for each layer reads
	\begin{align}
		H_{\rm intra}^{L}=&-A\sum_{\left\langle i,j\right\rangle}  {\bf m}_{i}^{L}\cdot{\bf m}_{j}^{L}+K\sum_{i}\left[ 1-({\bf m}_{i}^{L,z})^2\right] 
		\nonumber\\
		&+D_{ij}\sum_{\left\langle i,j\right\rangle}({\bf \nu}_{ij}\times\hat{z})\cdot({\bf m}_{i}^{L}\times{\bf m}_{j}^{L})+H_{\rm d}
		,\label{2}				
	\end{align}
	where $\left|  {\bf m}_{i}^{L}\right| =1$ represents the normalized local magnetic moment at site $i$, and $\left\langle i,j\right\rangle$ extends to all the nearest neighbor sites in each layer. The first term represents the FM Heisenberg exchange interaction with $A$ being the intralayer exchange constant. The second term represents the perpendicular magnetic anisotropy (PMA) with $K$ being the anisotropy constant. The third term represents the DMI, where $D_{ij}$ is the DMI constant and ${\bf \nu}_{ij}$ is the unit vector between sites $i$ and $j$. The fourth term $H_{\rm d}$ represents the dipole-dipole interaction. On the other hand, the interlayer Hamiltonian which describes the interaction between Co and Gd layers reads \cite{zhang2016thermally}
	\begin{align}
		H_{\rm inter}=&-\sigma c_z\sum_{i}  {\bf m}_{i}^{\rm Co}\cdot{\bf m}_{j}^{\rm Gd}
		,\label{3}				
	\end{align}
	where $c_z$ is the thickness of the cell size, and $\sigma$ is the bilinear surface exchange coefficient. 
	
	To explore the current-induced dynamics of FiM skyrmion, we numerically solve the Landau-Lifshitz-Gilbert (LLG) equation augmented with a spin-torque term $\tau$,
	\begin{align}
		\partial_t{\bf m}=-\gamma\,{\bf m} \times {\bf h}_{\rm eff}+ \alpha\left( {\bf m} \times \partial_t{\bf m}\right) +\tau
		,\label{4}		
	\end{align} 
	where $\gamma$ is the gyromagnetic ratio, $\alpha$ is the Gilbert damping coefficient, and $ {\bf h}_{\rm eff}=-(\partial H / \partial {\bf m}) /(\mu_0 M_{\rm s})$ is the effective local field with $\mu_0$ being the the vacuum permeability constant. We consider two strategies for the injection of spin-polarized currents \cite{tomasello2014strategy}. For the current-in-plane (CIP) geometry, the current flows through all the layers where the spin-transfer torques (STTs) are formulated in the Zhang-Li form \cite{zhang2004roles}:
	\begin{align}
		\tau_{\rm STT}=u\left( {\bf m} \times \partial_x{\bf m}\times {\bf m}\right) -\beta u\left( {\bf m} \times \partial_x{\bf m}\right) 
		.\label{5}		
	\end{align} 
	Here, the first term is the adiabatic torque, and the second term is the nonadiabatic torque with $\beta$ being the degree of nonadiabaticity. The STT coefficient is given by $u=(g\mu_BPJ)/(2eM_{\rm s})$ \cite{yamane2016spin}, where $g$, $\mu_B$, $P$, $e$ and $J$ are the Land\`{e} factor, Bohr magneton,   spin polarization factor, elementary charge and applied current density, respectively. On the contrary, for the current-perpendicular-to-plane (CPP) geometry, the charge current flows through the Pt layer and leads to an out-of-plane spin current due to the spin Hall effect \cite{tomasello2014strategy}. In this case, damping-like spin-orbit torque (SOT) works, which is formulated as \cite{slonczewski1996current}:
	\begin{align}
		\tau_{\rm SOT}=(\frac{g\mu_B \theta_{\rm H} J}{2ec_zM_{\rm s}})({\bf m} \times{\bf m}_p \times {\bf m})
		,\label{6}		
	\end{align}
	where $\theta_{\rm H}$ is the spin Hall angle, and ${\bf m}_p$ is the normalized spin polarization vector. Here, the field-like SOT is excluded as its impact on the skyrmion dynamics is negligible \cite{xia2019current}.
	
	The simulations are performed by MuMax3 finite-difference GPU accelerated code \cite{vansteenkiste2014design}. The simulation volumes for static and dynamic cases are $100\times100\times1\ {\rm nm}^3$ and $300\times200\times1\ {\rm nm}^3$, respectively, with a cell size of $0.5\ {\rm nm}$, matching the lattice constant of Co to maximize the calculation accuracy. The default simulation parameters and their ranges of variation are given in Table~\ref{tab:table1}.
	
	\begin{table}[b]
		\caption{\label{tab:table1}
		Simulation parameters.
		}
		\begin{ruledtabular}
			\begin{tabular}{ccccc}
				\textrm{Parameters}&
				\textrm{Units}&
				\textrm{Co layer}&
				\textrm{Gd layer}&
				\textrm{Refs.}\\
				\colrule
				$M_{\rm s}\footnote{For simplify, $M_{\rm s}^{\rm Gd}$ is always set to an estimated value of 50\% of $M_{\rm s}^{\rm Co}$ \cite{mullermodelling}.}$&${\rm MA}/{\rm m}$&1.44&0.72&\cite{mullermodelling,lalieu2017deterministic}\\
				$A$&${\rm pJ}/{\rm m}$&\multicolumn{2}{c}{15 [5, 20]\footnote{Values enclosed in square brackets denote the ranges of variation.}}&\cite{das2023bilayer}\\
				$K$&${\rm MJ}/{\rm m}^3$&2.0 [1.5, 3.0]&0&\cite{wang2018theory}\\
				$D$&${\rm mJ}/{\rm m}^2$&2.5 [1.5, 3.0]&0&\cite{das2023bilayer}\\
				$\sigma$&${\rm mJ}/{\rm m}^2$&\multicolumn{2}{c}{-10}&\cite{zhang2016magnetic}\\
				$\alpha$&-&\multicolumn{2}{c}{0.02}&\cite{jiang2017direct,berges2022size}\\
				$g\footnote{For simplify, an average value of $g=2.1$ is employed for both layers.}$&-&2.2&2.0&\cite{berges2022size}\\
				$J$&${\rm GA}/{\rm m}^2$&\multicolumn{2}{c}{1 [0.1, 10]}&\cite{zhang2016magnetic}\\
				$P$&-&\multicolumn{2}{c}{0.4}&\cite{xia2019current}\\
				$\theta_{\rm H}$&-&\multicolumn{2}{c}{0.05}&\cite{morota2011indication}\\
				$\varDelta M_{\rm s}$&-&\multicolumn{2}{c}{[-0.05, 0.05]}&-\\
			\end{tabular}
		\end{ruledtabular}
	\end{table}
	
	\begin{figure*}[t]
		\includegraphics[width=1\linewidth]{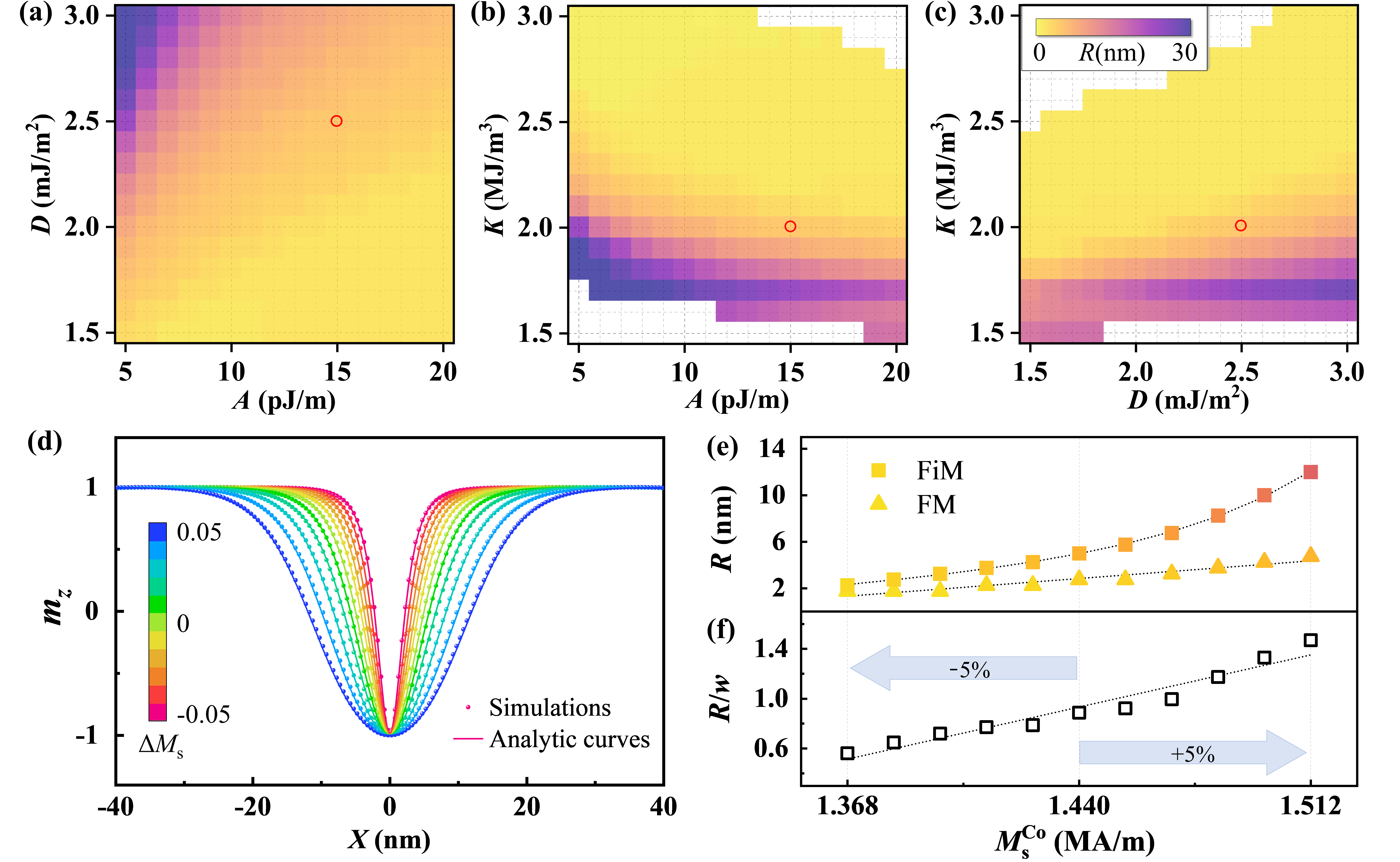}
		\caption{(a)-(c) Variations of equilibrium skyrmion radius $R$ versus material parameters $A$, $D$, and $K$. The color represents the skyrmion size, and the white region indicates the absence of a stable skyrmion. The red circles mark the situation with default parameters, i.e., $K=2.0$, $D=2.5$, and $A=15$. (d) Magnetization component $m_z$ plotted as a function of position $X$ along the skyrmion diameter under different $M_{\rm s}$. The symbols are simulation results, and the solid lines are fitted curves. (e) Variations of skyrmion radius $R$ for both FM and FiM cases versus $M_{\rm s}$. (f) Variations of the ratio of skyrmion radius to domain wall width $R/w$ versus $M_{\rm s}$. The symbols are simulation or analytical results, and the dashed lines are used to guide the eyes.
		}\label{2} 
	\end{figure*}
	
	\section{\label{sec:3}Results and discussion}
	\subsection{\label{sec:31}Static FiM skyrmion without a $\bf{\textit{M}}_{\bf s}$ gradient}
	We begin by verifying the configuration of a static FiM skyrmion in a system without a ${M}_{\bf s}$ gradient. For this purpose, we utilize a square geometry system of $100\times100\ {\rm nm}^2$ with infinite boundaries. A bilayer skyrmion is initialized in the system and the equilibrium state is then determined using the conjugate gradient method. Firstly, while maintaining the default $M_{\rm s}$, we investigate the impact of variation in material parameters $A$, $D$, and $K$ on the stability and size of the skyrmion. The results are depicted in Figs.~\ref{2} (a)-(c), where in each figure two parameters are varied with the remaining parameter being fixed at its default value. Overall, the skyrmion radius $R$ varies from a few nanometers to approximately 30 nm within the selected range, which is positively correlated with $D$, while negatively correlated with $A$ and $K$. When skyrmion size becomes excessively large or small, the system tends to change into multi-domain or uniform states. It is noteworthy that the stability of skyrmion does not necessarily require such a large PMA because similar configurations can be obtained by applying a vertical external magnetic field \cite{beg2015ground}. Based on the above results, we set $A=15$, $D=2.5$ and $K=2.0$ as default values, resulting in an initial skyrmion radius of 5.5 nm.
	
	Secondly, we fix $A$, $D$ and $K$ at their default values and vary $M_{\rm s}$ within a small range of $\pm5\%$ (i.e., $M_{\rm s}^{\rm Co}$ ranging from 1.386 to 1.512 while $M_{\rm s}^{\rm Gd}$ maintains half of the $M_{\rm s}^{\rm Co}$ value) \cite{mullermodelling} to observe variation in the skyrmion configuration. The $m_z$ profiles along the diameter direction under different $M_{\rm s}$ are depicted in Fig.~\ref{2} (d). The dots are extracted from simulation data, while the solid curves are results of fitting with \cite{wang2018theory}
	\begin{align}
		m_z(X)=2\tan^{-1}\left[\frac{\sinh (R/w)}{\sinh (X/w)}\right] 
		,\label{7}		
	\end{align}
	where $w$ is the width of the 360\textdegree domain wall of the skyrmion, considered as a fitting parameter, and $R/w$ is the normalized skyrmion radius. Given that our simulation results agree well with the theoretical formula, we thus extract $R$ and $R/w$ from Fig.~\ref{2} (d) and plot their variations with respect to $M_{\rm s}$ in Figs.~\ref{2} (e) and (f). For comparison, calculated values of $R$ in a single-layer FM are also plotted, where a notably higher sensitivity of FiM skyrmion size to $M_{\rm s}$ is observed. Note that $R/w$ also increases with increasing $M_{\rm s}$, and approximately exhibits a linear relationship, which has an important meaning in the subsequent theoretical analysis (see Sec. \ref{sec:33}).

	\subsection{\label{sec:32}Suppression of the skyrmion Hall effect}
	
	We first consider the case with constant $M_{\rm s}$. As described in Sec. \ref{sec:2}, both CPP and CIP injection geometries are initially considered for current-driven skyrmion motion. For CIP injection, we further examine two cases with different nonadiabatic degrees, $\beta=0$ and $\beta=\alpha$. The dependence of skyrmion velocity $v=\rVert{\bf v}\rVert$ on the current density $J$ for different scenarios is shown in Fig.~\ref{3} (a). The overlapping symbols indicate that skyrmions in the two layers always move as bounded entities. This means that the interlayer surface exchange $\sigma$ is sufficiently strong, thereby preventing their decoupling under high current. Moreover, the CPP injection turns out to be of much greater efficiency in driving skyrmions compared to the CIP injection, which is consistent with reported findings on FM skyrmions \cite{tomasello2014strategy} and AFM skyrmions \cite{zhang2016magnetic}. Therefore, in the following sections, we only consider the SOT-induced skyrmion motion for the sake of simplicity, with a moderate density of $1\ \rm{GA}/{\rm m}^2$.
	
	\begin{figure}[t]
		\includegraphics[width=1\linewidth]{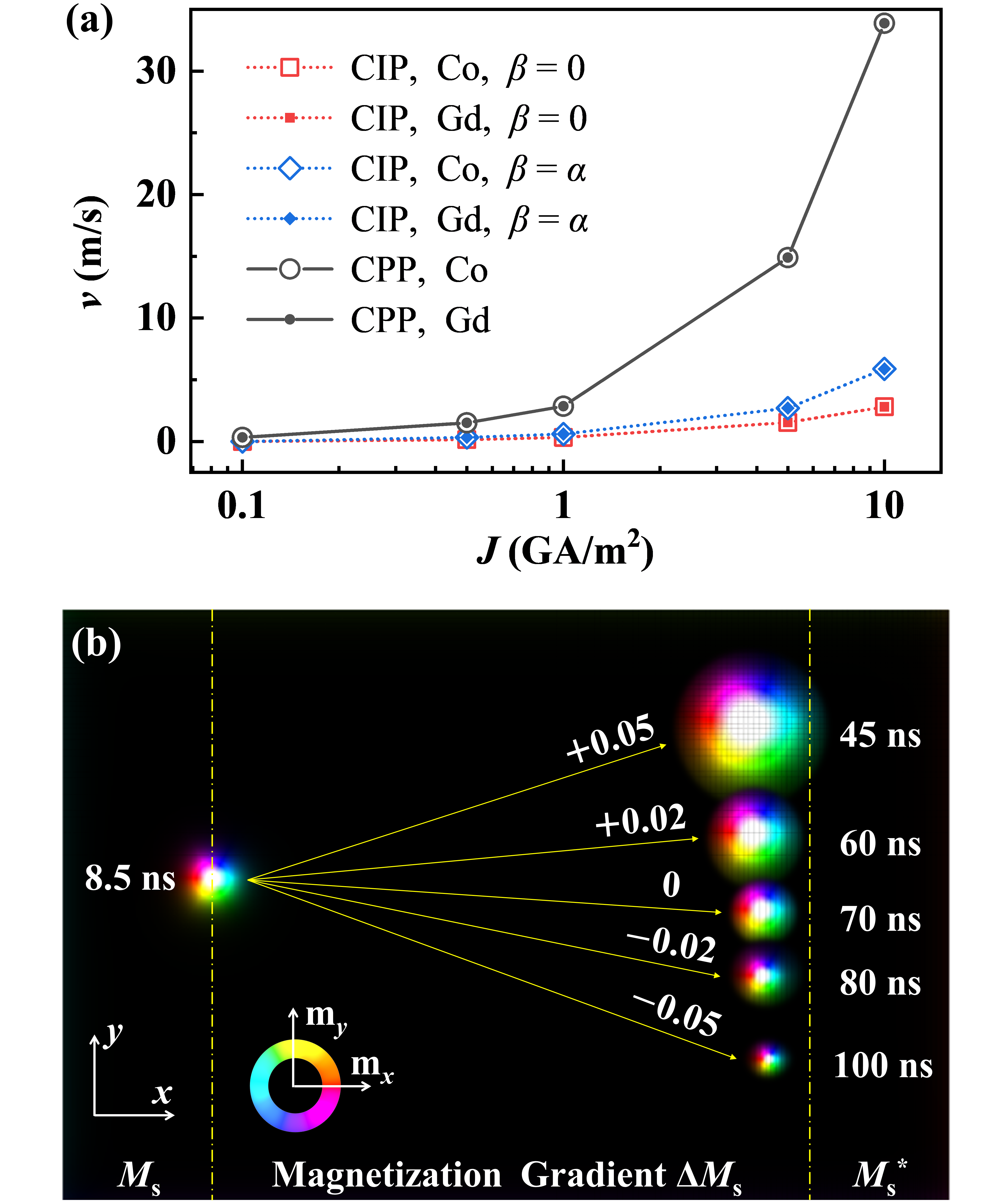}
		\caption{(a) Skyrmion velocity $v$ as functions of the driving current density $J$ for both CPP and CIP injections. The Co and Gd layers are represented by hollow and solid symbols, respectively. (b) Snapshots of the SOT-induced skyrmion dynamics for various $\varDelta M_{\rm s}$. The dashed lines separate regions with constant $M_{\rm s}$ or finite $M_{\rm s}$ gradient, and the arrows indicate the direction of skyrmion motion. The color wheel indicates the in-plane magnetization. $\varDelta M_{\rm s}$ and time information are indicated in white text.}\label{3}  
	\end{figure}
	
	Next, we introduce the $M_{\rm s}$ gradient in the central region of the nanoplate. The linear $\varDelta M_{\rm s}$ is defined as
	\begin{align}
		\varDelta M_{\rm s}=(M_{\rm s}^*-M_{\rm s})/M_{\rm s}
		,\label{8}		
	\end{align}
	where $\varDelta M_{\rm s}$ varies between $\pm0.05$ and is controlled by changing $M_{\rm s}^*$ at the right-side region. For various values of $\varDelta M_{\rm s}$, snapshots of the driven skyrmions near the left and right markers (yellow chain lines) of the gradient region are presented in Fig.~\ref{3} (b). At 8.5 ns, the skyrmion reaches the left marker and begins to move along different trajectories under the influence of different $\varDelta M_{\rm s}$. We find significant variations in the skyrmion size and velocity during the moving process. Positive $\varDelta M_{\rm s}$ leads to an expansion of the skyrmion and a decrease in velocity, while negative $\varDelta M_{\rm s}$ results in its contraction and an increase in velocity. The trajectories of the skyrmion present a scattered distribution. When $\varDelta M_{\rm s}=0$, there is a small but nonzero skyrmion Hall angle $\varPhi$, consistent with a previous report \cite{wang2022topological}. As $\varDelta M_{\rm s}$ varies, the trajectory exhibits a monotonic and regular change, providing visual evidence that the magnetization gradient can regulates the skyrmion Hall angle $\varPhi$. 
	
	To learn more details of the skyrmion motion under the $M_{\rm s}$ gradient, we further extract the data of skyrmion dynamics, which display the variations in skyrmion position $Y(X)$, radius $R$, and velocity components $v_x$ and $v_y$ as functions of the position $X$ in Fig.~\ref{4}. Here, to provide more precise outcomes, we only display the cases where $\varDelta M_{\rm s}$ varies from $-0.02$ to $+0.03$, within the central gradient range. The skyrmion center is determined by averaging the position coordinates of the 360\textdegree\ domain wall with $m_z=0$. The results in Fig.~\ref{4} (a) are similar to those in Fig.~\ref{3} (b), but we can observe that the trajectories of the skyrmion center are remarkably smooth. It is noteworthy that there is almost no displacement in the $y$-direction  when $\varDelta M_{\rm s}=+0.01$. Figure~\ref{4} (b) reveals linear dependence between $R$ and $X$ with a slope determined by $\varDelta M_{\rm s}$, where positive (negative) $\varDelta M_{\rm s}$ tends to induce the expansion (shrinkage) of the skyrmion. Importantly, the dynamically changing values of $R$ agree well with those in the equilibrium states in Fig.~\ref{2} (e). This agreement is expected when the skyrmion velocity is relatively low because the driven skyrmion can be relaxed to approach the minimal-energy configuration at equilibrium at each moment. Therefore, we believe that the findings about the static configuration in Fig.~\ref{2} (e) are also applicable for the analysis of the driven skyrmions. The velocity components, $v_x$ and  $v_y$, are crucial parameters in determining $\varPhi$. Hence, we plot them separately in Fig.~\ref{4} (c). When $\varDelta M_{\rm s}$ is positive (negative), both $v_x$ and $v_y$ increase (decrease) as compared with those at $\varDelta M_{\rm s}=0$. The difference appears in their spatial-position dependence, that is, $v_y$ is nearly independent of the position, whereas $v_x$ exhibits nearly linear dependence on $X$.
	
	\begin{figure}[t]
		\includegraphics[width=1\linewidth]{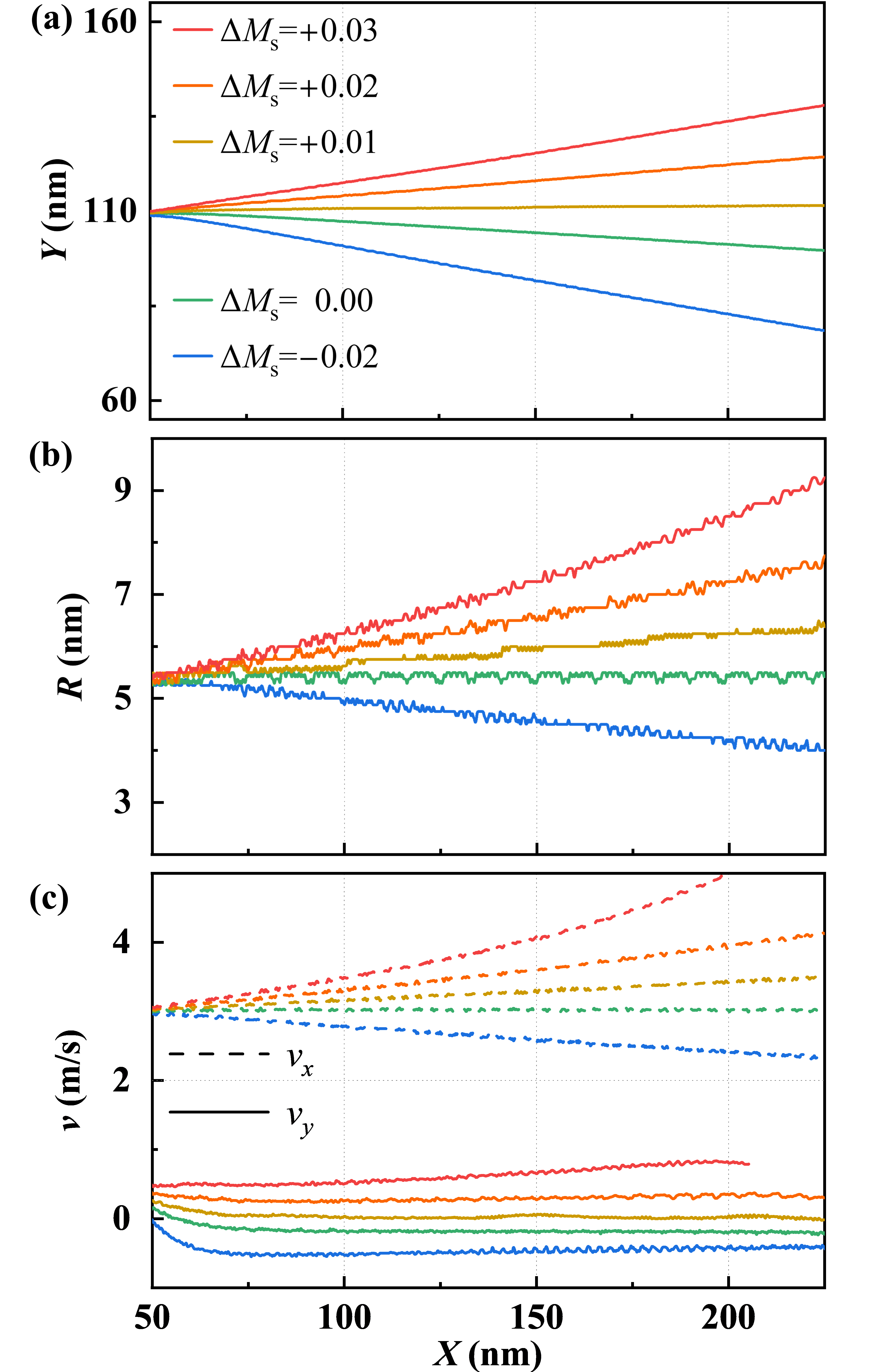}
		\caption{ (a) Variations of the skyrmion center coordinates $(X, Y)$ in the central region during the motion process. (b) Variations of the skyrmion radius $R$ versus position $X$ for various $\varDelta M_{\rm s}$. (c) Variations of the skyrmion velocity components, $v_x$ and $v_y$, versus position $X$. The dashed lines represent $v_x$, and the solid lines represent $v_y$. The color of the line indicates different $\varDelta M_{\rm s}$.
		}\label{4} 
	\end{figure}
	
	After obtaining the skyrmion instantaneous velocity ${\bf v}(v_x,v_y)$, we can then calculate the skyrmion Hall angle $\varPhi$ at each moment by
	\begin{align}
		\varPhi=\tan^{-1} (v_y/v_x)
		.\label{9}
	\end{align}
	As the motion of skyrmion is a continuous process, we are more interested in its total displacement over a certain period of time. Therefore, we further calculate the average value of $\varPhi$ throughout the entire motion, presenting it alongside the instantaneous values in Fig.~\ref{5} (a). Figure~\ref{5} (a) is a violin plot with the instantaneous and average values of $\varPhi$ being represented by hollow and solid symbols, respectively. The curves on both sides of the violin shape display the probability density distribution, which clearly indicate a Gaussian distribution for all $\varDelta M_{\rm s}$. When the variation in $\varDelta M_{\rm s}$ is within 5\% (ranging from $-0.02$ to $+0.03$), the change in $\varPhi$ can reach up to $\pm10$\textdegree, and it exhibits an approximately linear correlation with $\varDelta M_{\rm s}$. Specifically, when $\varDelta M_{\rm s}$ is $+0.01$, the average value of $\varPhi$ is only $0.06$\textdegree, indicating an extremely small SkHE.
	
	\begin{figure}[t]
		\includegraphics[width=1\linewidth]{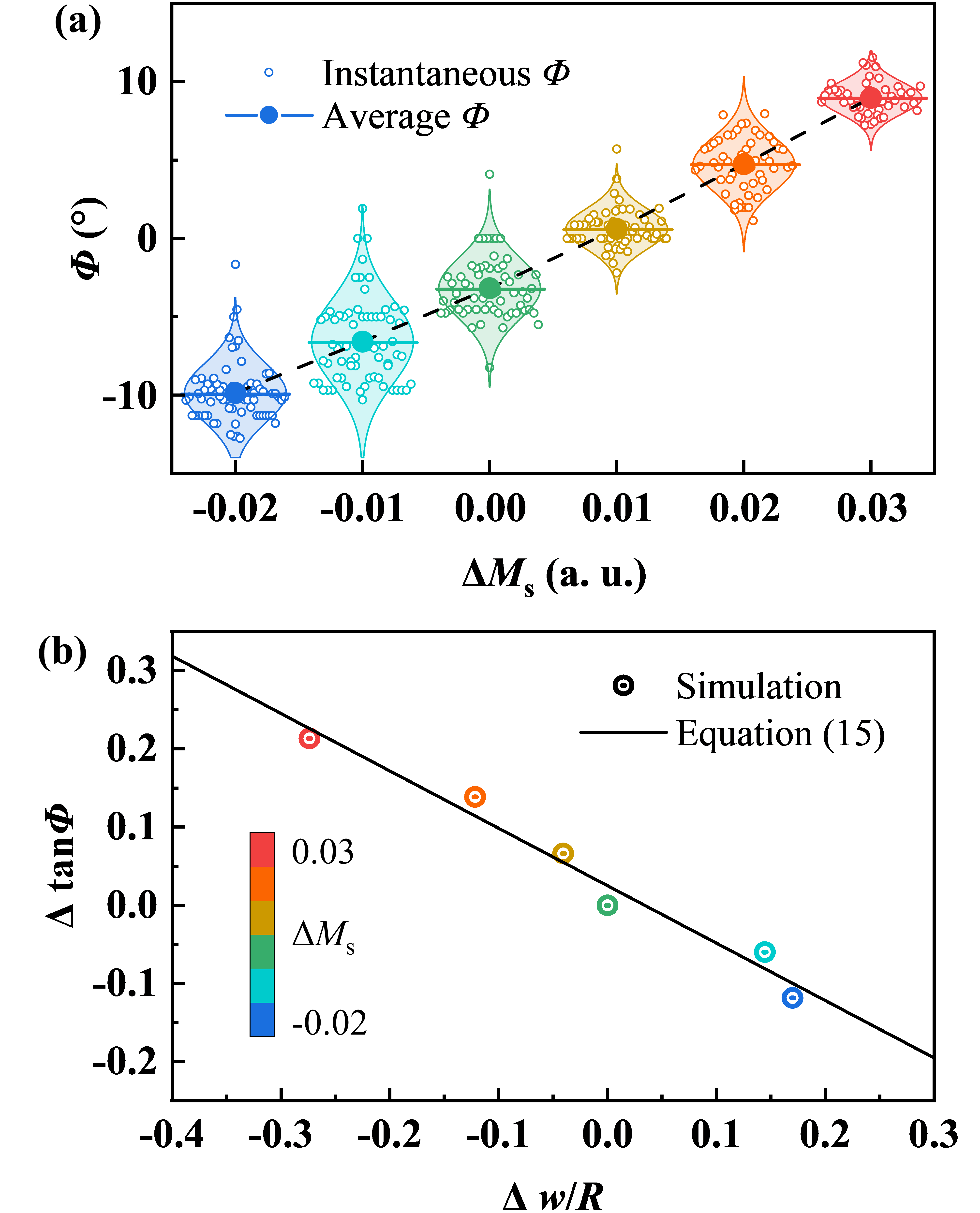}
		\caption{ (a) Violin plot with hollow and solid symbols depicting the instantaneous and average skyrmion Hall angles $\varPhi$ for various $\varDelta M_{\rm s}$. The side-profile lines represent the probability density distribution of instantaneous values. The dashed line illustrates the variation of $\varPhi$ versus $\varDelta M_{\rm s}$. (b) The increment of skyrmion Hall angle $\varDelta \tan \varPhi$ plotted as a function of the increment of the reciprocal of the normalized skyrmion radius $\varDelta w/R$. The symbols represent simulation results, with colors indicating various $\varDelta M_{\rm s}$. The solid line is a linear fitted line based on Eq. (\ref{15}).
	}\label{5} 
	\end{figure}
	
	\subsection{\label{sec:33}Analysis based on Thiele equation}
	In this section, we theoretically analyze of the simulation results. While the size of a N\'{e}el skyrmion may change, its central symmetry in shape remains unchanged, which allows us to model it as a rigid point-like particle. Therefore, its motion can be qualitatively elucidated by a modified Thiele equation \cite{thiele1973steady}
	\begin{align}
		{\bf G} \times {\bf v} -\alpha \mathcal{D} \cdot {\bf v} + \mathcal{B}\cdot{\bf J}= \bf{0}
		.\label{10}
	\end{align}
	The three terms arise from the precessional, damping, and SOT
	terms in the LLG equation, respectively. Here, $ {\bf G}=n\hat{z} $ is the gyromagnetic coupling vector, $\mathcal{D}$ is the dissipative tensor, and the tensor $\mathcal{B}$ quantifies the efficiency of the spin Hall torque over the two-dimensional configuration. The parameters $n$ and $d$ correspond respectively to the topological invariant as a sum of the solid angles of magnetizations, and the component of the disspation tensor as an indicator of the magnetization rotation length-scale, given by 
	\begin{align}
	n=\iint{\bf m}\cdot \left({\partial_x {\bf m}} \times {\partial_y {\bf m}}\right) \, d^2r
	,\label{11} \\
	d=\mathcal{D}_{xx}=\mathcal{D}_{yy}=\iint({\partial_x {\bf m}})^2 \, d^2r
	.\label{12} 
	\end{align}
	For a FM skyrmion with a simple profile, $d$ is approximately given by $d\approx \pi^3R/w$ \cite{jiang2017direct}, while for a FiM skyrmion with $R$ roughly as large as $w$, it is given by $d\approx{2\pi R}/{w}$ \cite{berges2022size}. This means that $d\propto R/w$ is expected to scale likewise with the normalized skyrmion radius $R/w$ presented in Fig.~\ref{2} (e). In our model, due to the presence of spatially dependent $\varDelta M_{\rm s}$, the skyrmion internal configuration $R/w$ changes with $X$, resulting in the variation of $d=kR/w$ during the skyrmion motion, where $k$ is the proportionality coefficient.
	
	The solutions of Eq. (\ref{10}) are
	\begin{align}
		v_x&=B_0J\left( \frac{\alpha d}{1+\alpha^2d^2}\right)
		,\label{13} \\
		v_y&=B_0J\left( \frac{n}{1+\alpha^2d^2}\right)
		,\label{14} 
	\end{align}
	where $B_0$ is a constant derived from the tensor $\mathcal{B}$. Equations (\ref{13}) and (\ref{14}) can be used to explain the results of instantaneous velocity in Figs.~\ref{4} (c) and (d). When $\alpha$ is small enough, the term $\alpha^2d^2$ can be neglected, so that $v_y$ becomes independent of $d$, manifested in Fig.~\ref{4} (d) as $v_y$ almost unrelated to $X$. As for $v_x$, the term $\alpha d$ cannot be neglected, so that $v_x$ is related to $d$, i.e., correlated with $R/w$, as seen in Fig.~\ref{4} (c), where $v_x$ exhibits an approximate linear correlation with $X$.
	
	Regarding the variation of the skyrmion Hall angle $\varPhi$, it follows the relation
	\begin{align}
		\tan \varPhi\equiv \frac{v_y}{v_x}=\frac{kn}{\alpha}\cdot\frac{w}{R}
		.\label{15}  
	\end{align}
	The above Eq. (\ref{15}) can be used to explain the simulation results shown in Fig.~\ref{5} (a). The original variable $\varDelta M_{\rm s}$ affects the internal skyrmion configuration $R/w$, causing an incremental change $\varDelta d$ during the motion process. This further leads to an increment of $\varDelta \varPhi$. When $\varDelta M_{\rm s}$ takes an appropriate value, around $+0.01$ in this work, $\varDelta \varPhi$ precisely compensates for the original $\varPhi$, thus suppressing the skyrmion Hall effect. To visually demonstrate the relationship in Eq. (\ref{15}), we plot the function of $\varDelta \tan \varPhi$ with respect to $\varDelta w/R$, as shown in Fig.~\ref{5} (b). This function reveals the correlation between the changes in $\varPhi$ and the normalized skyrmion radius. We find that the simulation results for different $\varDelta M_{\rm s}$ are uniformly distributed around the theoretical line, and the linear relationship conforms to the description provided by Eq. (\ref{15}).

	\section{\label{sec:4}Conclusion}
	In conclusion, we have investigated the static properties and current-induced dynamics of the synthetic FiM skyrmions. Compared to the FM skyrmions, equilibrium FiM skyrmions exhibit a higher sensitivity of their static profiles to the variation in $M_{\rm s}$. Inspired by the graded-$M_{\rm s}$ in the field of graded-index magnonics, we have explore the influence of a $M_{\rm s}$ gradient on the skyrmion dynamics by particularly focus on the SOT-induced motion. It is shown that the presence of a $M_{\rm s}$ gradient significantly affects the trajectory, velocity, and size of the driven skyrmion. Notedly, a 5\%\ variation in $M_{\rm s}$ gradient can control the skyrmion Hall angle within a range of $\pm10$\textdegree. Specifically, a 1\%\ $M_{\rm s}$ gradient completely suppresses the SkHE. We have analyzed the simulation results using Thiele theory, demonstrating that the $M_{\rm s}$ gradient regulates the skyrmion Hall angle by influencing the normalized skyrmion radius. In this study, we have incorporated the $M_{\rm s}$ gradient from the magnonics to the skyrmionics. Our findings are expected to provide a new perspective to overcome the SkHE, which has been an obstacle to the technical applications of skyrmions so far.

	\begin{acknowledgments}
		X.F.Z. acknowledges support by the National Science Fund for Distinguished Young Scholars (Grant No. 52225312), the National Natural Science Foundation of China (Grant No. U1908220), and the Key Research and Development Program of Zhejiang Province (Grant No. 2021C01033). X.C.Z. and M.M. acknowledge support by the JST CREST (Grant No. JPMJCR20T1). M.M. also acknowledges support by the JSPS KAKENHI (Grants No. JP20H00337, No. JP23H04522 and No. JP24H02231), and the Waseda University Grant for Special Research Projects (Grant No. 2024C-153). L.B. acknowledges the financial support by China Scholarship Council (Grant No. 202206080023).
	\end{acknowledgments}

    \bibliography{Manuscript}
	
\end{document}